\documentclass[prl,aps,preprint,superscriptaddress]{revtex4-1}
\usepackage{marvosym}
\usepackage{graphicx,subfigure}         %For figures
\usepackage{amsmath} 
\usepackage{amssymb,stmaryrd}   %For mathematical equations
\usepackage{color}
\usepackage{wrapfig}
\usepackage{mathtools}
\usepackage{tikz}  
\usetikzlibrary{arrows,shapes,trees,positioning}  
\usepackage{comment} 
\usepackage{units}
\usepackage{siunitx}

\graphicspath{{./exp_figs/}{./theory_figs/}}

\begin{document}
\title{\textbf{Ultrafast magnetization dynamics in half-metallic Co$_{2}$FeAl Heusler alloy}}

\author{R. S. Malik }%\email[]{rameezsaeed.malik@physics.uu.se}
\affiliation{Department of Physics and Astronomy, Uppsala University, Box $516$, SE-$751 20$, Uppsala, Sweden}

\author{E. K. Delczeg-Czirjak}%\email[]{erna.delczeg@physics.uu.se}
\affiliation{Department of Physics and Astronomy, Uppsala University, Box $516$, SE-$751 20$, Uppsala, Sweden}

\author{D. Thonig}%\email[]{danny.thonig@ou.se}
\affiliation{School of Science and Technology, \"Orebro University, SE-$701 82$ \"Orebro, Sweden}

\author{R. Knut}
\affiliation{Department of Physics and Astronomy, Uppsala University, Box $516$, SE-$751 20$, Uppsala, Sweden}

\author{I. Vaskivskyi}
\affiliation{Department of Physics and Astronomy, Uppsala University, Box $516$, SE-$751 20$, Uppsala, Sweden}

\author{R. Gupta}
\affiliation{Department of Materials Science and Engineering, Uppsala University, Box $534$, SE-$751 21$, Uppsala, Sweden}

\author{S. Jana}
\affiliation{Department of Physics and Astronomy, Uppsala University, Box $516$, SE-$751 20$, Uppsala, Sweden}

\author{R. Stefanuik}
\affiliation{Department of Physics and Astronomy, Uppsala University, Box $516$, SE-$751 20$, Uppsala, Sweden}

\author{Y. O. Kvashnin}
\affiliation{Department of Physics and Astronomy, Uppsala University, Box $516$, SE-$751 20$, Uppsala, Sweden}

\author{S. Husain}
\affiliation{Department of Materials Science and Engineering, Uppsala University, Box $534$, SE-$751 21$, Uppsala, Sweden}

\author{A. Kumar}
\affiliation{Department of Materials Science and Engineering, Uppsala University, Box $534$, SE-$751 21$, Uppsala, Sweden}

\author{P. Svedlindh}
\affiliation{Department of Materials Science and Engineering, Uppsala University, Box $534$, SE-$751 21$, Uppsala, Sweden}

\author{J. S\"oderstr\"om}
\affiliation{Department of Physics and Astronomy, Uppsala University, Box $516$, SE-$751 20$, Uppsala, Sweden}

\author{O. Eriksson}
\affiliation{Department of Physics and Astronomy, Uppsala University, Box $516$, SE-$751 20$, Uppsala, Sweden}
\affiliation{School of Science and Technology, \"Orebro University, SE-$701 82$ \"Orebro, Sweden}

\author{O. Karis}
\affiliation{Department of Physics and Astronomy, Uppsala University, Box $516$, SE-$751 20$, Uppsala, Sweden}

\begin{abstract}
% Half-metallic Heusler alloys
%, with high Curie temperature, 
% have 100\% spin polarization at Fermi level ($E_{F}$), which makes them suitable materials for spintronics devices. 
We report  on optically induced, ultrafast magnetization dynamics in  the Heusler alloy $\mathrm{Co_{2}FeAl}$, probed by
%thin films 
time-resolved magneto-optical Kerr effect. Experimental results are compared to results from electronic structure theory and atomistic spin-dynamics simulations. 
Experimentally, we find that the demagnetization time ($\tau_{M}$) in films of $\mathrm{Co_{2}FeAl}$ is almost independent of varying structural order, and that it is similar to that in elemental 3d ferromagnets. 
%It is proposed that electronic mechanisms govern this part of the optically induced magnetization dynamics. 
In contrast, the slower process of magnetization recovery, specified by $\tau_{R}$, is found to occur on picosecond time scales, and  is demonstrated to correlate strongly with the 
%theoretically calculated values of the 
Gilbert damping parameter ($\alpha$). Our results show that  $\mathrm{Co_{2}FeAl}$ is unique, in that it is the first material that clearly demonstrates the importance of the damping parameter in the remagnetization process. Based on these results we argue that for $\mathrm{Co_{2}FeAl}$ the remagnetization process is dominated by magnon dynamics, something which might have general applicability. 
%The theoretical calculations demonstrate strong ferromagnetic Heisenberg interactions that result in local,  Weiss-type, fields that are almost independent on degree of structural disorder. In addition, theory shows a monotonous decrease of the damping parameter, $\alpha$, as the structure evolves from $\textit{A}$2 to $\textit{B}$2 to $L2_{1}$ order. 
%This decrease in damping reflects the half-metallicity in CFA, as the density of states (DOS) decrease at $E_{F}$ with increase in $\textit{B}$2 ordering.

\end{abstract}  

\maketitle
%\section{Introduction}

 %1. what is ultrafast demagnetization and why we are interested in ultrafast demagnetization (general).

Studies of ultrafast demagnetization was pioneered by Beaurepaire \textit{et al.}\cite{beaurepaire1996}, who demonstrated that the optical excitation of a ferromagnetic material - using a short pulsed laser could quench the magnetic moment on  sub-picosecond timescales. The exact underlying microscopic mechanisms responsible for the transfer of angular momentum have been strongly debated for more than 20 years \cite{scholl1997,koopmans2005unifying,koopmans2010explaining}. Ultrafast laser-induced demagnetization has now become an intense field of research not only from fundamental point of view but also from a technological aspect, due to an appealing possibility to further push the limits of operation of  information storage and data processing devices \cite{kirilyuk2016}. Both experiment \cite{cinchetti2006spin,stamm2007femtosecond,dalla2007influence,walowski2008energy,carpene2008dynamics,koopmans2010explaining,roth2012temperature,mathias2012probing,rudolf2012ultrafast,eschenlohr2013ultrafast,turgut2013controlling,kirilyuk2016} and theory \cite{zhang2000laser,koopmans2005microscopic,krauss2009ultrafast,steiauf2009elliott,bigot2009coherent,battiato2010superdiffusive,essert2011electron,atxitia2011ultrafast} report that all of the 3d ferromagnets (Fe, Ni and Co) and their alloys, show characteristic demagnetization times in the sub-picosecond range, while 4f metals exhibit a complicated two-step demagnetization up to several picoseconds after the excitation pulse \cite{koopmans2010explaining,radu2015ultrafast}.

% Why heusler alloys are interested in general and why specifically for ultrafast. what make them unique for ultrafast demagnetization studies

In this work, we have made element specific investigations of the ultrafast magnetization dynamics of  a half-metallic Heusler alloy. This class of alloys has been investigated intensively, especially concerning the magnetic properties, ever since the discovery in 1903, when Heusler \textit{et al.} reported that alloys like $\mathrm{Cu_{2}MnAl}$ exhibit ferromagnetic properties, even though none of its constituent elements was  in itself ferromagnetic  \cite{heusler1903}. The ferromagnetic properties were found to be related to the chemical ordering \cite{persson1929}. One of the key features of several Heusler alloys is their unique electronic structure, where the majority spin band-structure has a metallic character while the minority spin band is semiconducting with a band gap.
%These Heusler alloys have the peculiarity that the majority spin band-structure has a metallic character while minority spin band-structure has semiconducting character%
Such materials are also referred to as half-metallic ferromagnets (HMFs) and  were initially predicted by de Groot \textit{et al.} \cite{de1983new}, based on electronic structure theory. Half-metals ideally exhibit  100\% spin-polarization at the Fermi level. This exclusive property makes them candidates to be incorporated in spintronic devices, e.g.\ spin filters, tunnel junctions and giant magneto-resistance (GMR) devices \cite{spintronics2004,spinfilters2002,tsymbal2001GMR,palmstromheusler}. One of the advantages of Heusler alloys with respect to other half-metallic system, like $\mathrm{CrO_{2}}$ and $\mathrm{Fe_{3}O_{4}}$, are their relatively high Curie temperature ($T_{c}$) and low coercivity ($H_{c}$) \cite{alloys1991,heusler}. Heusler alloys are also appealing for spintronic applications due to the low Gilbert damping, which allows for a long magnon diffusion length \cite{brown2000,kobayashi2004, graf2011simple,shaw2018magnetic,liu2009origin,schoen2016ultra}. It has been shown that the low value of the Gilbert damping constant is related with the half-metallicity \cite{shaw2018magnetic,liu2009origin}. The origin of the band gap and the mechanism of half-metallicity in these materials have been studied by using first principle electronic structure calculations \cite{GalanakisPRB_01,galanakis2007spin,katsnelson2008,kandpal2007calculated}.
 % In a half metal, since only one spin-type is present at the Fermi level, it is possible to achieve high spin polarization by combining different elements to tune the Fermi level ($E_{F}$) into the middle of the band gap \cite{graf2011simple}.  
%The Co-based Heusler alloys with the chemical formula $\mathrm{Co_{2}XY}$ consist of four interpenetrating face-centered cubic (fcc) lattices (X= Cr, Mn, Fe; Y= Al, Ga, Si). 
%In addition, the Heusler alloys are of particular interest because of high spin-polarization and low damping.\cite{nakatani2007structure,brown2000,kobayashi2004,graf2011simple,shaw2018magnetic,liu2009origin,schoen2016ultra} 
The half-metallic property is furthermore known to be very sensitive to structural disorder
\cite{balke2008rational,galanakis2007spin,katsnelson2008,picozzi2004role,kandpal2007calculated}.
From a fundamental point of view, it is intriguing to ask, how the band gap in the minority spin channel effects the ultrafast magnetization dynamics of Heusler alloys \cite{muller2009spin,mann2012insights}. It has already been reported that some of the half-metals like $\mathrm{CrO_{2}}$ and $\mathrm{Fe_{3}O_{4}}$ exhibit very slow dynamics, involving time-scales of hundreds of picoseconds, \cite{mann2012insights,muller2009spin} while several Co-based Heusler alloys show a much faster demagnetization, similar to the time-scales of the elemental 3d-ferromagnets \cite{steil2014ultrafast,steil2010band,wustenberg2011ultrafast}.
%One of the possible reasons for a fast dynamics in some of these Heusler alloys, is the low degree of chemical ordering in the crystal structure. This is typically connected to a higher density of states (DOS) at the Fermi level, and a wanishing bandgap.\cite{picozzi2004role,ozdougan2011effect} 
The faster dynamics of these Heuslers has been discussed in Ref.\cite{muller2009spin} to be due to the fact that the band gap in the minority spin channel is typically around 
$\unit[0.3-0.5]{eV}$, which is smaller than the photon energy ($\unit[1.5]{eV}$) of the exciting laser. It is also smaller than the band gap of $\mathrm{CrO_{2}}$ and $\mathrm{Fe_{3}O_{4}}$. Importantly, the Heusler alloys offer the possibility to study magnetization dynamics, as a function of structural order, since they normally can be prepared to have a fully ordered $L2_{1}$ phase, a partially ordered $\textit{B}$2 phase, and a completely disordered $\textit{A}$2 phase. The structural relationships of these phases are described in the Supplemental Material (SM) \cite{SM}.
  
We have here studied the optically induced, ultrafast magnetization dynamics of $\mathrm{Co_{2}FeAl}$ (CFA) films, using time-resolved magneto-optical Kerr effect (TR-MOKE) as described in Ref.\ \cite{somnath_01}. By control of the growth temperature, CFA alloy forms with varying degree of structural order, in a continuous way between the $\textit{A}2$ and $\textit{B}2$ phases, as well as between the $\textit{B}2$ and $L2_{1}$ phases \cite{mizukami2009low,kumar2017temperature}. 
%
%As will be demonstrate here, this enables a deeper understanding of the microscopic mechanisms behind optically induced magnetization dynamics in this class of materials, as well as for magnetic materials in general.
We present data from four CFA samples, grown at $\unit[300]{K}$, $\unit[573]{K}$, $\unit[673]{K}$, and $\unit[773]{K}$ respectively. We henceforth denote each sample by its growth temperature as a subscript, e.g. $\mbox{CFA}_\mathrm{300 K}$. As evidenced by X-ray diffraction, the sample grown at $\unit[300]{K}$ is found to exhibit the A2 phase, while the samples grown at $\unit[573]{K}$ and $\unit[673]{K}$ predominantly exhibit the B2 phase. The sample grown at $\unit[773]{K}$ is found to exhibit a pure B2 phase \cite{kumar2017temperature}. The value of the Gilbert damping $\alpha$ is found to monotonously decrease with annealing temperature and is thus lowest for the sample grown at $\unit[773]{K}$
\cite{foot:commentonkumar}.  

%The CFA films prepared here, with different ordering, exhibit a %similar demagnetization time as in 3d transition metals. However, %remagnetization occurs on a longer timescale than for the %ferromagnetic transition metals. We have found that the %remagnetization time correlates with the Gilbert damping of these %films, which gives insights to the process of magnetization %dynamics in these systems.

%%%%%%%%%%%%%%%%%%%%%%%%%%%%%%%%%%%%%%%%%%%%%%%%%%%%%%%%%%%%%%%%%%%%%%%%%%%%
\begin{figure}[h!tbp]
	%\centering
	\includegraphics[width=1\columnwidth]{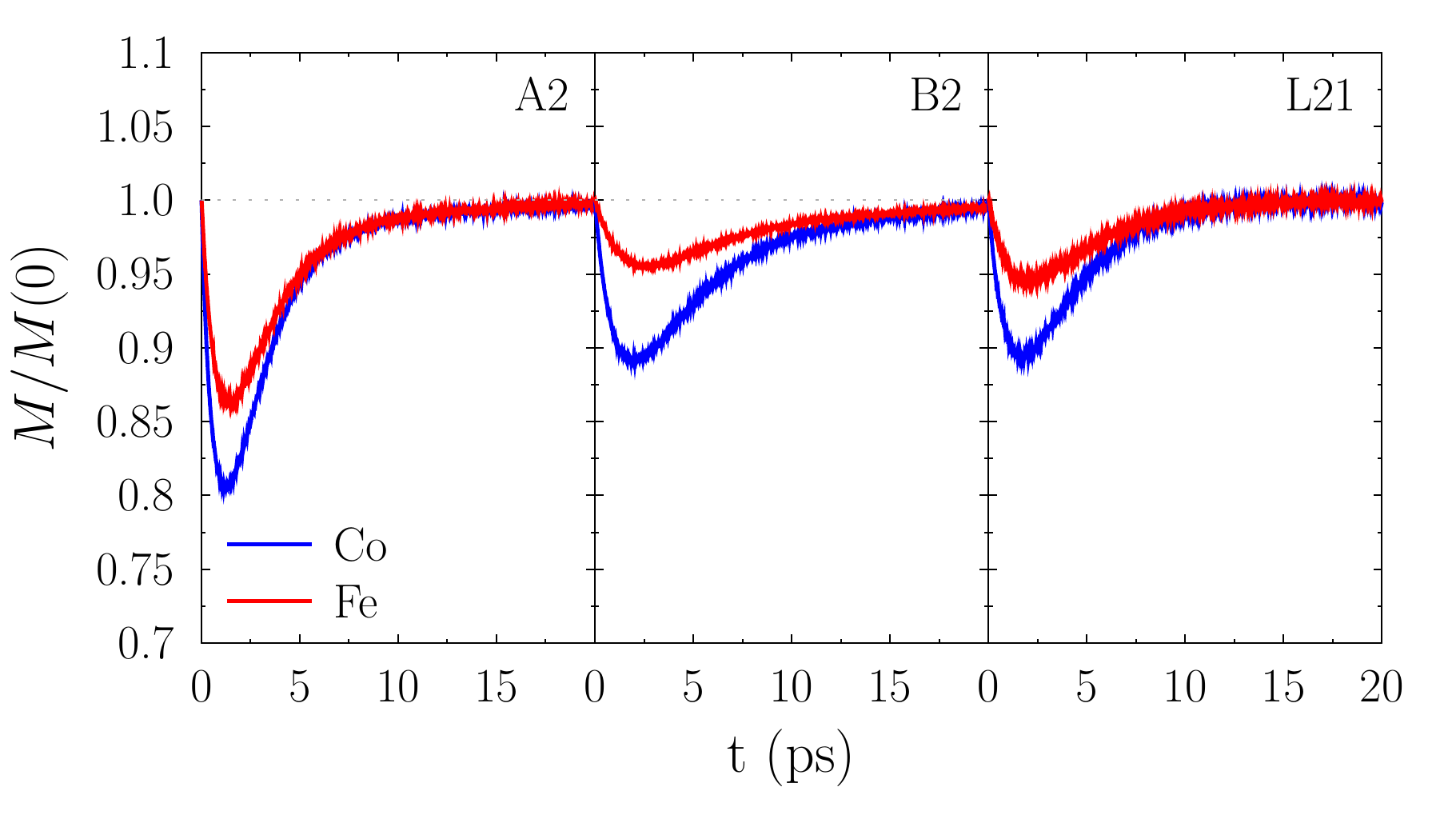}
	\caption{(Color online) Simulations of ultrafast dynamics of Co$_2$FeAl in the different structural phases $A2$ (left panel), $B2$ (central panel), and $L2_1$ (right panel). The demagnetization is shown element resolved (blue line - Co, red line - Fe). The peak temperature is $\unit[1200]{K}$. The dotted line indicates the equilibrium magnetization at $T=\unit[300]{K}$.}
	\label{fig:demagtheo}
\end{figure}

Calculations based on density functional theory (DFT) of the magnetic moment, Heisenberg exchange interaction and the Gilbert damping parameter are described in detail in (SM) \cite{SM}. These parameters were used in a multiscale approach to perform atomistic magnetization dynamics simulations, described in Sec.S1 of (SM) \cite{SM}. Here we employed the two temperature model (2TM) for the temperature profile of the spin-system.
%(see Fig.~\ref{fig:demagtheo}). 
In the 2TM, the spin temperature increases due to the coupling to the hot-electron bath, that is excited by the external laser pulse. In the simulations we used a peak temperature in the 2TM of $\unit[1200]{K}$. A full description of the 2TM and the details of all spin-dynamics simulations are described in Sec.S2 of SM \cite{SM}. 

The results of the simulations are shown in Fig.~\ref{fig:demagtheo}, for the $\textit{A}2$, $\textit{B}$2 and $L2_{1}$ phases. It can be seen that the different phases react differently to the external stimulus.
In general, this model provides a dynamics that is controlled by \textit{i)} the temperature of the spin-subsystem, \textit{ii)} the strength of the magnetic exchange interaction and \textit{iii)} the dissipation of angular momentum and energy during the relaxation of the atomic magnetic moments (Gilbert damping) \cite{Eriksson:2016uw}. 
%Consequently, the time evolution of the average magnetic moment is expected to be different for the different degrees or ordering, i.e. for $\textit{A}$2, $\textit{B}$2, and L2$_1$. 
% With this choice, we observe a reduction of the Co (Fe) moment of about $20\%$ ($17\%$) in the $\textit{A}$2, $12\%$ ($9\%$) in the $\textit{B}$2, and $12\%$ ($12\%$) in the L2$_1$ phase. 
Before continuing the discussion, we note that the average magnetization, $M$, of element $X$ is calculated as $M^X = \nicefrac{\sum_i c_i^X M^X_i}{4\sum_i c_i^X}$, where $c_i^X$ is the concentration of the particular element $X$ in the particular phase and $i$ runs over the four nonequivalent sites of the unit cell. After the material demagnetizes, the spin temperature eventually drops and the average magnetization returns to its initial value after $\unit[10-20]{ps}$ (cf. Fig.~\ref{fig:demagtheo}). 

\begin{figure}[tbhp]
	%\centering
	%\includegraphics[width=0.5\columnwidth]{theory_figs/relaxtimes_theory.pdf}
		\includegraphics[width=0.75\columnwidth]{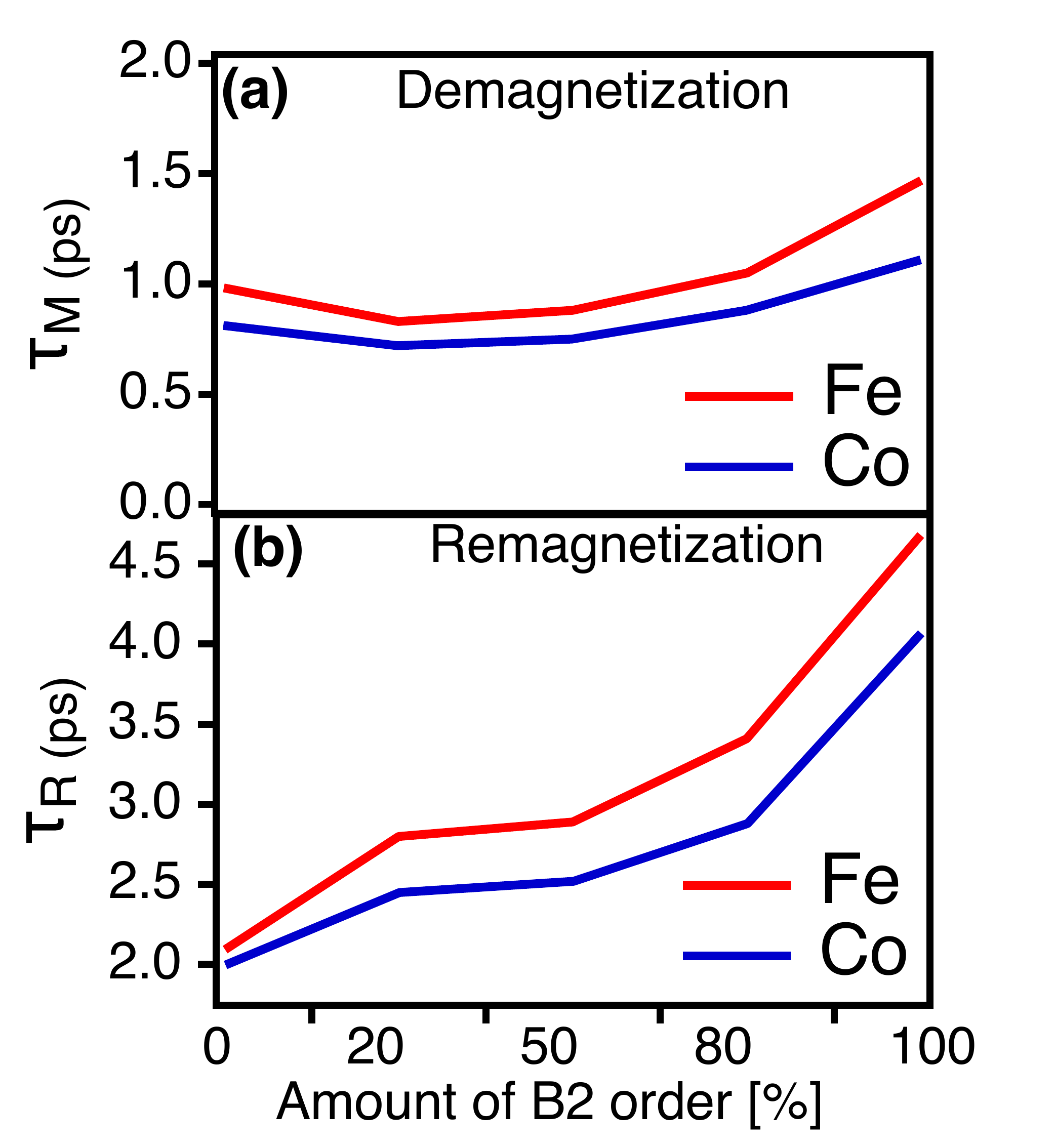}
	\caption{(Color online) Element resolved relaxation times of Co$_2$FeAl, from simulations of alloys with varying amounts of $A2 \to B2$ phase. 0 corresponds to pure $A2$ phase while 100 corresponds to pure $B2$ phase. Panel (a) shows the demagnetization time and panel (b) shows the remagnetization time. Both time constants are obtained from fitting the time trajectory of $M^X(t)$ by a double-exponential function (see text).}
	\label{fig:timestheo}
\end{figure}

To estimate the time constants of the demagnetization $\tau_{M}$ and remagnetization ($\tau_{R}$) processes, in an element-specific way, we fit both the theoretical and experimental transient magnetizations by a double exponential function \cite{atxitia2010evidence}.
%To estimate the timescales of the de- and remagnetization processes, in an element-specific way, we fit $M^X(t)$ by a double-exponential function as proposed in Ref.~\cite{koopmans2010explaining} and obtain from the simulations the demagnetization time $\tau_{M}$ and remagnetization time $\tau_{R}$.
We show results of $\tau_{M}$ and $\tau_{R}$ in  Fig.~\ref{fig:timestheo} for the $\textit{A}2$ and $\textit{B}$2 phase, as well as for alloys with intermediate degree of disorder (described in Sec.S2 of SM \cite{SM}). 
% In the theoretical calculations, focus was made for these alloys, since the experimental results are primarily for these phases. 

The theoretical demagnetization time is seen from Fig.~\ref{fig:timestheo} to typically be around $\unit[1]{ps}$, whereas the remagnetization time is $\unit[2-5]{ps}$. Going from the $\textit{A}$2 to the $\textit{B}$2 alloy, both times increase, albeit the simulations show a stronger increase of the remagnetization time as function of alloy composition. We also note that the relevant time scale is somewhat larger for Fe than for Co, and the ratio between them, $\nicefrac{\tau_{Fe}}{\tau_{Co}}$, grows when going from $\textit{A}$2 to $\textit{B}$2 phase. 
%The trends illustrated in Fig.~\ref{fig:timestheo} follow the behavior of the inverse of the Gilbert damping and will be discussed in more detail below. 
%arameter for the magnetisation dynamics. 
%his was also discussed in Ref.~\onlinecite{chimata2017magnetism} 
%{\bf (Rameez please add here a ref. to: R. Chimata et al, Phys. Rev. B 95, 214417 (2017) )} RSM:Fixed
%nd given that all other interactions are the same, one could expect that $\tau\approx \frac{1}{\alpha}$. %However, it should be noticed that the %demagnetization times $\tau^{demag}$ can be %different from experimental measurements. 

%\begin{itemize}
%    \item Optical transitions mediated by dipole operator
%    \item Maybe Tc's
%\end{itemize}

%%%%%%%%%%%%%%%%%%%%%%%%%%%%%%%%%%%%%%%%%%%%%%%%%%%%%%%%%%%%%%%%%%%%
%\subsection{Ultrafast dynamics measurements}\label{sec:results}

\begin{figure*}[tbph]
            \includegraphics[width=\textwidth]{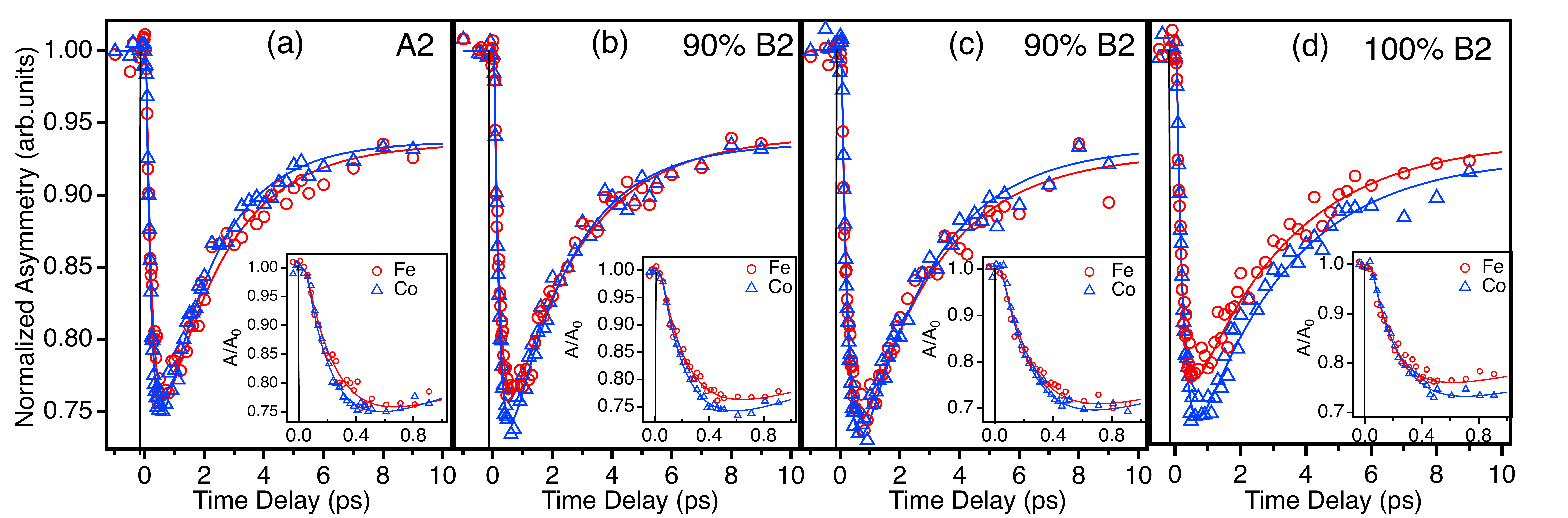}
	\caption{Measured element-specific Fe (red) and Co (blue) magnetization dynamics of Co$_2$FeAl. Samples are denoted by the growth temperature in each case. The red and blue lines correspond to fitted data (see text). (a) $\unit[300]{K}$ (100\% $\textit{A}_2$ phase), 
	(b) $\unit[573]{K}$ (90\% $B2$ phase), 
	(c) $\unit[673]{K}$ (90\% $B2$ phase), and 
	(d) $\unit[773]{K}$ (100\% $B2$ phase). 
	The insets show the demagnetization dynamics up to $\sim$ $\unit[1]{ps}$. All of the measurements were performed with similar pump-fluence (for details, see Sec.S4 of SM).}
\label{fig:CFA_dynamics}
\end{figure*}

%In this section, we describe results from our measurements and the influence of chemical ordering on the ultrafast magnetization dynamics in the CFA samples.  
Figures \ref{fig:CFA_dynamics} (a-d) shows the measured magnetization dynamics of CFA films that were grown at different temperatures (see SM, Sec.S3 for thin films synthesis, and Sec.S4 for details on the experimental measurements \cite{SM}). The inset shows the  observed magnetization dynamics up to $\sim$1 ps. For all  samples, the data for Fe (red) and Co (blue)  show  similar demagnetization dynamics in the first few hundred femtoseconds, whereupon differences in the magnetization dynamics become visible, especially on the picosecond timescale. 
%The experimental data is fitted using the analytical solution of the three temperature model, and relevant parameters are extracted, such as $\tau_{M}$ and $\tau_{R}$ \cite{atxitia2010evidence}.
 
% {\bf here we need a reference}. %In all samples, the demagnetization time $\tau_{M}$ for Co and Fe is same. 

\begin{figure}
		%\centering\includegraphics[width=0.55\columnwidth]{exp_figs/CFA_demag_exp.pdf}
	    %\centering\includegraphics[width=0.55\columnwidth]{exp_figs/CFA_remag_exp.pdf}
	    \centering\includegraphics[width=0.75\columnwidth]{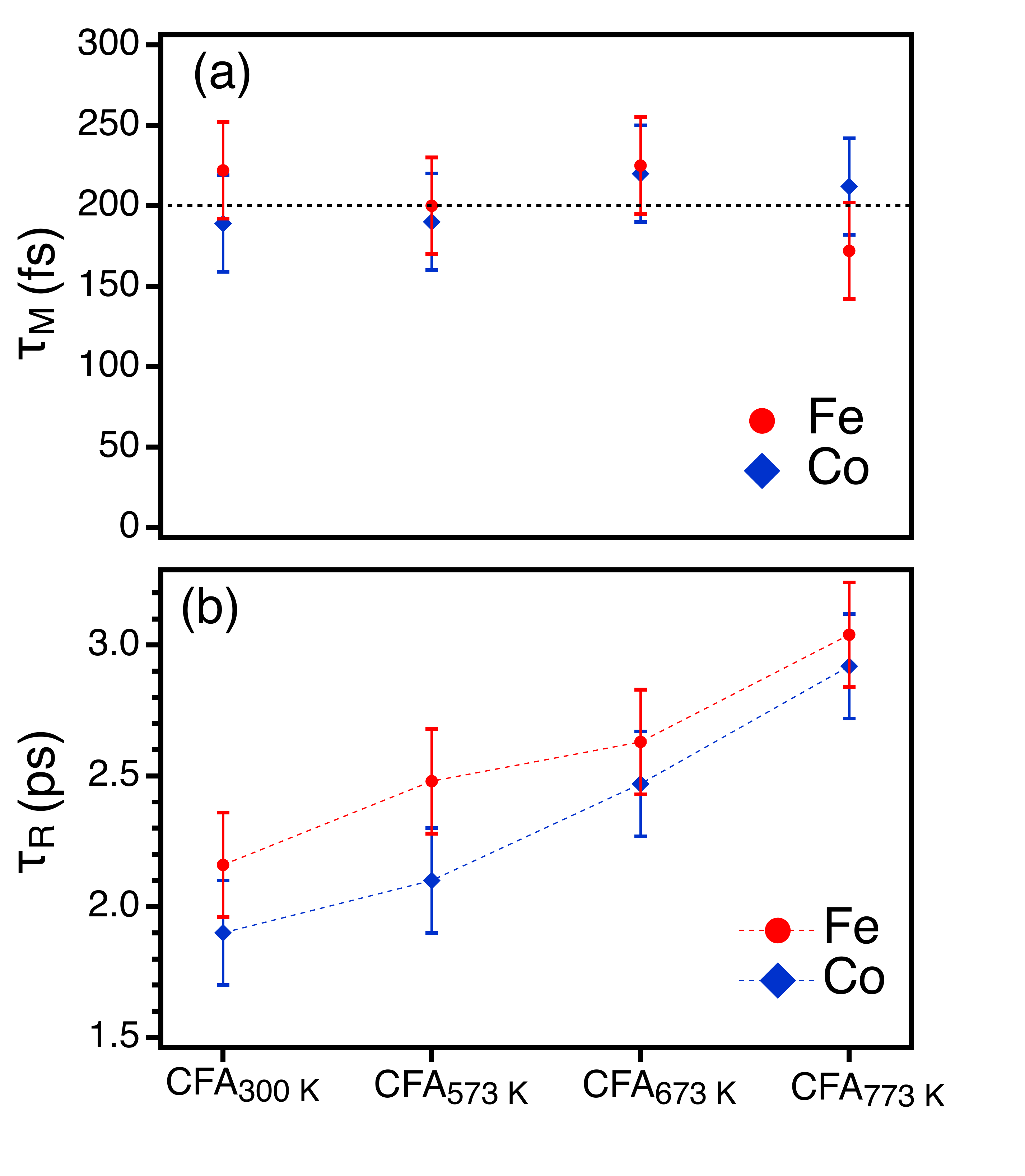}
\caption{Measured magnetization times for the investigated Co$_2$FeAl alloys. In (a) the demagnetization time, $\tau_{M}$, is shown and in (b) the remagnetization time, $\tau_{R}$, is plotted.}
\label{fig:demag_remag}
\end{figure}

Figure \ref{fig:demag_remag} (a-b) shows the measured values of the demagnetization and remagnetization time constants, for the four different growth temperatures, representing different degree of disorder in Co$_2$FeAl, along the alloy path $\textit{A}2 \to \textit{B}$2. It may be seen that the $\tau_{M}$ for Fe and Co is the same within the error bars for all four samples, regardless of the degree of structural ordering (Fig.~\ref{fig:demag_remag}a). It may also be noted that the measured $\tau_{M}$ for CFA is similar to that of 3d transition metals \cite{steil2010band,steil2014ultrafast} and very much shorter than that of $\mathrm{CrO_{2}}$ or $\mathrm{Fe_{3}O_{4}}$.

Demagnetization times that are independent on degree of structural ordering is interesting, since it can be expected that the presence of structural disorder in Heusler alloys ought to result in a lower degree of spin polarization of the electronic states (i.e.\ an increased density of states (DOS) at the Fermi level in the minority band). This is expected to enhance spin-flip scattering,
%and the damping parameter in general, 
with an accompanying speed-up of the demagnetization dynamics \cite{muller2009spin,mann2012insights}. The electronic structure calculation of CFA also shows that the DOS at the Fermi level varies with different structural phases (analyzed in the Sec.S1 of SM \cite{SM}). The \textit{A}2 phase has a large number of states at the Fermi level, while the $L2_{1}$ phase, and to some extent the \textit{B}2 phase, has a low amount \cite{kumar2017temperature}. Despite these differences in the electronic structure, the measured demagnetization dynamics shown in Fig.~\ref{fig:demag_remag}(a) is essentially independent on degree of structural ordering.
%which correlates with the damping parameter %\cite{schoen2016ultra}.
%{\bf Olles question: do we need the section below?}
%The $\textit{A}$2 phase has large DOS at the Fermi level ($E_F$) and the $\textit{B}$2 phase has a smaller DOS.%whereas the $L2_{1}$ phase has a perfect half-metallic structure, having a gap in the minority spin channel.In addition, the value of DOS is low for this phase.Given such marked differences in the electronic structure, one might expect distinctly different magnetization dynamics, on all time-scales,%These large number of DOS at Fermi level enhanced the %spin-flip scattering process and one would expect a faster %demagnetization time ($\tau_{M}$)
%The demagnetisation time ($\tau_{M}$) for this sample is %expected to be longer as compare to A2 phase. and a longest %($\tau_{M}$) is expected among all other phases.
%Surprisingly, as Fig.~7a shows,  the demagnetization dynamics for all samples is similar within the experimental error bars and does not depend on the degree of structural ordering. 
%The inset of Fig. \ref{fig:CFA_dynamics} shows this dynamics for both Fe and Co. 
%A similar demagnetization dynamics was also observed for  $\mathrm{Co_{2}MnSi}$ and $\mathrm{Co_{2}FeSi}$ samples regardless of the different chemical ordering \cite{steil2014ultrafast,steil2010band}. 

On longer time-scales, there is a significant effect of structural ordering on the observed magnetization dynamics, which becomes particularly relevant for the remagnetization process. As seen in Fig.~\ref{fig:demag_remag}b, there is a monotonous increase of remagnetization time, $\tau_{R}$, with increasing growth temperature and hence the degree of ordering along the $\textit{A}2 \to \textit{B}$2 path. 
%As the ordering is improved in these sample,  $\tau_{R}$  also increases. %The remagnetization dynamics of Fe ad Co in $\mbox{CFA}_{(300 K)}$ is fast which give rise to a shorter $\tau_{R}$ time. The $\mbox{CFA}_{(773 K)}$ shows the slower remagnetization dynamics for both Fe and Co and a large $\tau_{R}$ time constant.
The sample grown at $\unit[300]{K}$ with $\textit{A}2$ phase, exhibits the fastest remagnetization dynamics ($\tau_{R}$). With increasing growth temperature and corresponding  increase in the structural ordering along the $\textit{A}$2 $\rightarrow$ $\textit{B}$2 path, a distinct trend of increasingly slower remagnetization dynamics is observed.

 %The $\mbox{CFA}_{(773 K)}$ sample has 100 \% $\textit{B}$2 phase and shows the slowest remagnetization time ($\tau_{R}$) among all other samples.
%Spin-polarized DOS calculation in Fig.~\ref{fig:dos} show that %the density of states varies in different phases of %$\mathrm{Co_{2}FeAl}$, as the structure ordering improve from %$\textit{A}$2 to $\textit{B}$2 and a band gap is formed for %perfectly order $L2_{1}$ phase.
%Interestingly in all sample, Co shows a faster remagnetization dynamics than Fe.

\begin{figure}
	\centering
	\includegraphics[width=1\columnwidth]{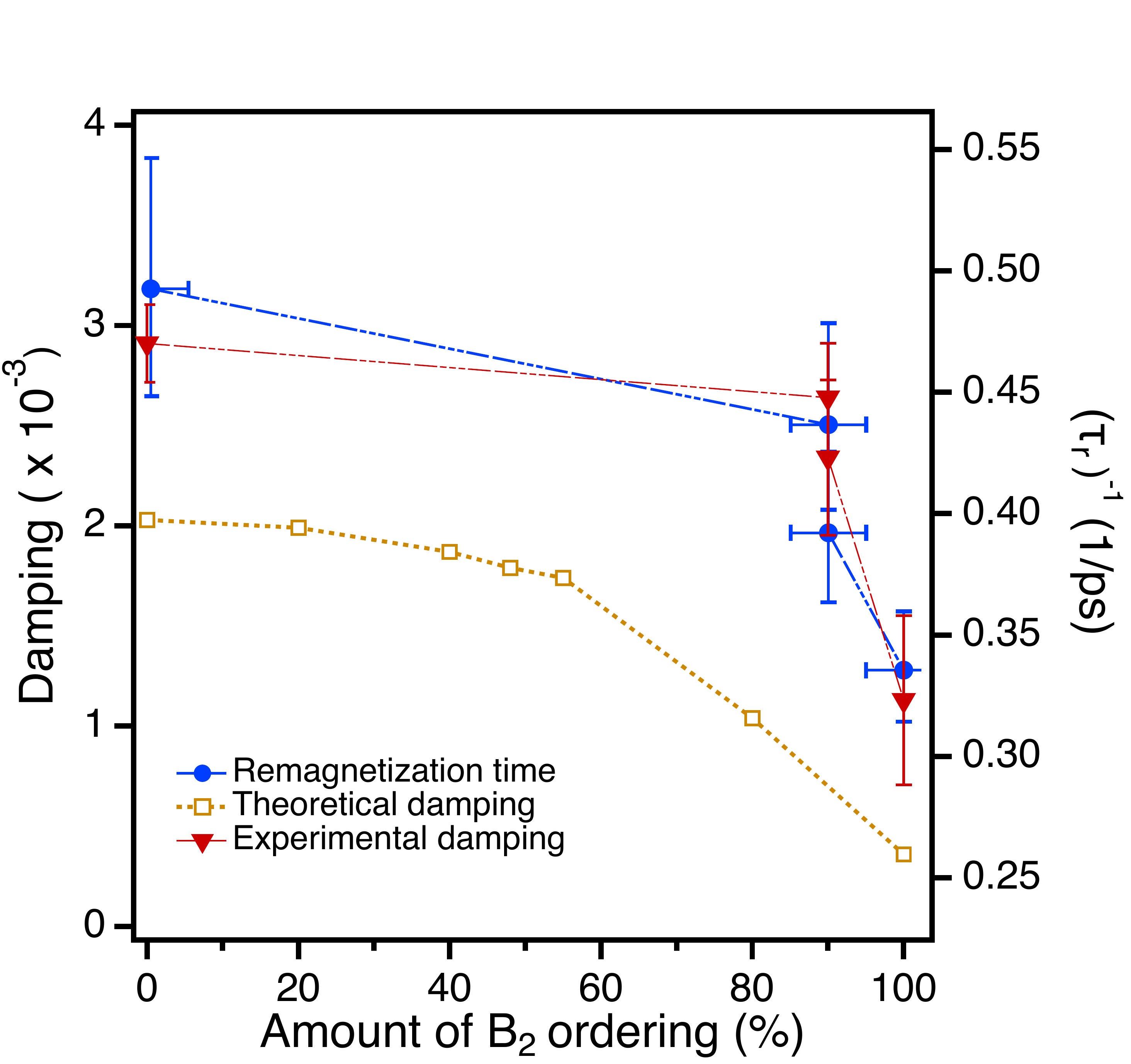}
	\caption{The relationship of inverse of the measured remagnetization time (right y-axis) and theoretically calculated and experimentally measured Gilbert damping (left y-axis) in Co$_2$FeAl for varying amount of B2 order  along the $\textit{A}2 \to \textit{B}$2 path, i.e.\ 0 corresponds to pure A2 phase while 100 corresponds to pure B2 phase. %Note that the theoretical values of the damping have been shifted upward with 0.001, since the calculations seem to make a slight underestimation of the damping parameter of these alloys%
	}
	\label{fig:damping_remag}
\end{figure}

The most conspicuous behaviour of the measured magnetization dynamics, and its dependence on the degree of ordering, concerns the remagnetization time (Fig.~\ref{fig:demag_remag}b).
%The relationship between Gilbert damping ($\alpha$) and %magnetization dynamics, especially for time-scales where the %dynamics is expected to be governed by magnons is highlighted in %Fig.~\ref{fig:damping_remag}. Here, we compare the theoretically %calculated Gilbert damping for Fe and Co in different phases of %$\mathrm{Co_{2}FeAl}$ with their remagnetization time $\tau_{R}$.
 %{\bf (note that we need to define how we get numbers on the x-axis here. Erna can you do this ? I think we should have it in the section on computational details. ED: added)}
%In contrast to Fig.~\ref{fig:Weissfield}, 
%The data in Fig.~\ref{fig:damping_remag} shows an interesting correlation. 
The time-scale of the remagnetization process is sufficiently long to allow for an interpretation based on atomistic spin-dynamics. Two materials specific parameters should be the most relevant to control this dynamics; the exchange interaction, as revealed by the local Weiss field, and the damping parameter. In the Sec.S2 of SM \cite{SM}, we report on the calculated Weiss fields and damping parameters. It is clear from these results that the trend in the experimental data shown in Fig.~\ref{fig:demag_remag}b, can not be understood from the Weiss field alone, whereas an explanation based on the damping is more likely. In order to illustrate this, we show in 
Fig.~\ref{fig:damping_remag} the inverse of the measured remagnetizatiom time compared to the theoretically calculated damping and experimental measured damping through ferromagnetic resonance (FMR) (described in Sec.S6 of SM)\cite{SM}.
The figure shows that the damping is large in the completely disordered $\textit{A}2$ phase and for a large range of structural orderings, which comes out from both theory and experiment.
%(in the perfectly ordered phase ($L2_{1}$) the damping is much smaller \cite{kumar2017temperature,mizukami2009low}).
The figure also demonstrates that the inverse of the measured remagnetization time scales very well with both the calculated damping and experimentally measured damping . According to the figure, a large damping parameter corresponds to faster remagnetization dynamics in the measurements.
%If magnons primarily govern the remagnetization process, and if the strength of the local - atomic Weiss field is similar for the different phases (see Fig. 6 in Appendix), Eqn. 3 suggests that the Gilbert damping is the most natural mechanism for angular momentum and energy transfer in these experiments. 

%\section{Discussion and Conclusion}\label{sec:discussion}

$\mathrm{Co_{2}FeAl}$ is, to the best of our knowledge, the first system where experimental observations and theory point to the importance of damping in the process of ultrafast magnetization dynamics. We note that this primarily is relevant for the remagnetization process; the initial part of the magnetization dynamics (first few hundred fs) is distinctly different. In the demagnetization we observe a similar behaviour for Fe and Co in all samples, and an insensitivity of the demagnetization times in relation to structural ordering. Also, the measured and theoretical demagnetization times evaluated from atomistic spin-dynamics simulations, do not agree. Other mechanisms, of electronic origin, most likely play role in this temporal regime.

The remagnetization process of $\mathrm{Co_{2}FeAl}$ alloys with varying degree of structural order, highlights clearly the importance of the Gilbert damping and that magnon dynamics dominates the magnetization at ps time-scales. The relevance of the Gilbert damping parameter for ps dynamics is natural, since this controls angular momentum (and energy) transfer to the surrounding. What is surprising with $\mathrm{Co_{2}FeAl}$ is the fact that other interactions (e.g. the Weiss field) show such a weak dependence on the amount of structural diorder. This is fortuitous, since it allows to identify the importance of the Gilbert damping.
A picture emerges from the results presented here, that the magnetization dynamics in general have two regimes; one which is primarily governed by electronic processes, and is mainly active in the first few hundered fs ($\tau_{M}$), and a second regime where it is primarily magnons that govern the remagnetiztion dynamics ($\tau_{R}$).

\begin{acknowledgments}
We acknowledge support from the Swedish Research Council (VR, contracts  2019-03666,2017–03799, 2016-04524 and 2013-08316), the Swedish Foundation for Strategic Research, project “SSF Magnetic materials for green energy technology” under Grant No. EM16-0039, the Knut and Alice Wallenberg foundation, STandUP and eSSENCE, for financial support. The Swedish National Infrastructure for Computing (SNIC) is acknowledged for computational resources.
\end{acknowledgments}
\bibliographystyle{apsrev4-2}
\bibliography{CFA_ref}
\end{document}